\RequirePackage{fix-cm}
\documentclass[smallextended]{svjour3}       
\smartqed  
\usepackage{graphicx}
\usepackage{subfigure}
\begin{document}

\title{Towards slime mould electrical logic gates with optical coupling}


\author{Richard Mayne         \and
        Andrew Adamatzky 
}


\institute{R. Mayne and A. Adamatzky \at
              Unconventional Computing Centre, University of the West of England, 
              Bristol, UK\\
              \email{Richard2.Mayne@live.uwe.ac.uk}            \\
              \email{Andrew.Adamatzky@uwe.ac.uk}
           }

\date{Received: date / Accepted: date}

\maketitle

\begin{abstract}

\emph{Physarum polycephalum} is a macroscopic single celled plasmodial slime mould. We employ plasmodial phototactic responses to construct laboratory prototypes of NOT and NAND logical gates with electrical inputs/outputs and optical coupling; the slime mould plays dual roles of computing device and electrical conductor. Slime mould logical gates are fault tolerant and resettable. The results presented here advance our understanding of how biological computing substrates may be manipulated to implement logical operations and demonstrate the feasibility of integrating living substrates into silicon hardware. 
\end{abstract}

\section{Introduction}

	Biological computing devices are currently a source of intense interest in both research and industry: massively-parallel, amorphous computing substrates such as slime moulds or mammalian brains are far superior to conventional solid-state computing architectures in a trade-off between energy consuming, self-growth and  processing ability, and our rapid approach to the physical limitations of inorganic material, the comparative rarity of earth metals used in electronic components and the toxic by-products produced in solid-state computer manufacture are all prime reasons to investigate unconventional computing devices. Furthermore, if sufficiently developed, such devices may allow us insight into other novel bio-computing technologies, e.g. neurally-integrated medical or augmentative prosthetic implants. \emph{P. polycephalum} is an ideal unconventional computing substrate as it is easy to culture, tolerant to abuse and has no ethical issues surrounding its use~\cite{Keller_Everhart_2010}.

The computing potential of the myxomycete slime mould \emph{Physarum polycephalum's} plasmodium (vegetative life-cycle stage, pl. plasmodia) has become a hot topic in computer science, biology and biophysics. Behaving as an excitable or a reaction-diffusion system encapsulated in an elastic membrane~\cite{Adamatzky_2007}, the slime mould may approximate shortest paths between two points~\cite{Nakagaki_2001} and solve maze problems \cite{Adamatzky_2012a,Nakagaki_2001}. Slime mould can be regarded as a proliferating massively parallel processor (a 'Physarum machine' \cite{Adamatzky_2010a}), capable of simultaneous processing of inputs (light, chemical gradients, temperature etc.) \cite{Adamatzky_2007}, concurrent decision making and distributed actuation. In experimental situations, slime mould based computing devices employ chemotaxis --- positive, towards chemoattractants, or negative, away from repellents --- or photoavoidance; the plasmoidum is motile via the contraction of muscle-like proteins invoking rhythmic shuttle streaming of its hydrodynamic core \cite{Adamatzky_2012b,Ishikawa_etal_1991}.

Previously we have demonstrated how \emph{P. polycephalum} may be hybridised with nano-scale metallic particles in order to manually alter the electrical properties of the plasmodium, e.g. facilitate the conduction of electricity and generate permanent, solid-state wires which persist following the plasmodium's migration and/or death \cite{Mayne_etal_2013,Mayne_Adamatzky_2013}.

In this study, we explore the use of light as a programmable input into a Physarum machine, and use the results to implement a basic yet functional device. Significant progress has been made programming a Physarum machine with spatially distributed nutrient sources~\cite{Nakagaki_2001,Adamatzky_2012b,Adamatzky_2010a}, but these are undynamic and problematic to `reprogram' (i.e. alter the input mid-experiment). Indeed, viable logic gates have been designed and produced by a number of authors using different variations of chemical attractant and repellent models~\cite{Tsuda_Aono_Gunji_2004,Adamatzky_2010b,Jones_Adamatzky_2010}, but all suffer from the aforementioned restrictions. These limitations may potentially be overcome by utilising light as an input, or rather a coupling medium between the sub-circuits of logical gates.

Two logic gates were designed: a NOT gate (more commonly known as an inverter), chosen for being the simplest logic operation, and a NAND gate, for being a universal gate, i.e. any logical operation can be carried out by multiple NAND gates.

\section{Selecting inputs: Phototaxis experiments}
\label{phototaxis}

\begin{figure}[!tbp]
\centering
\subfigure[]{\includegraphics[width=0.6\textwidth]{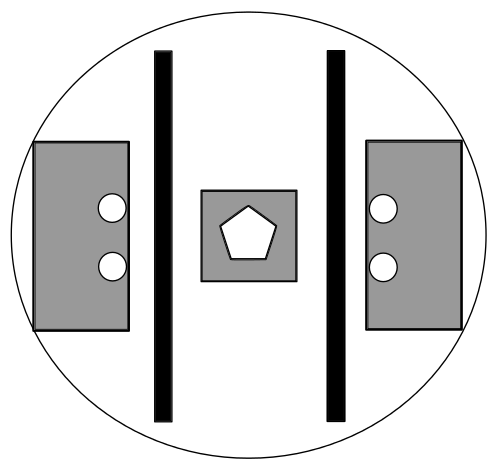}}
\subfigure[]{\includegraphics[width=0.9\textwidth]{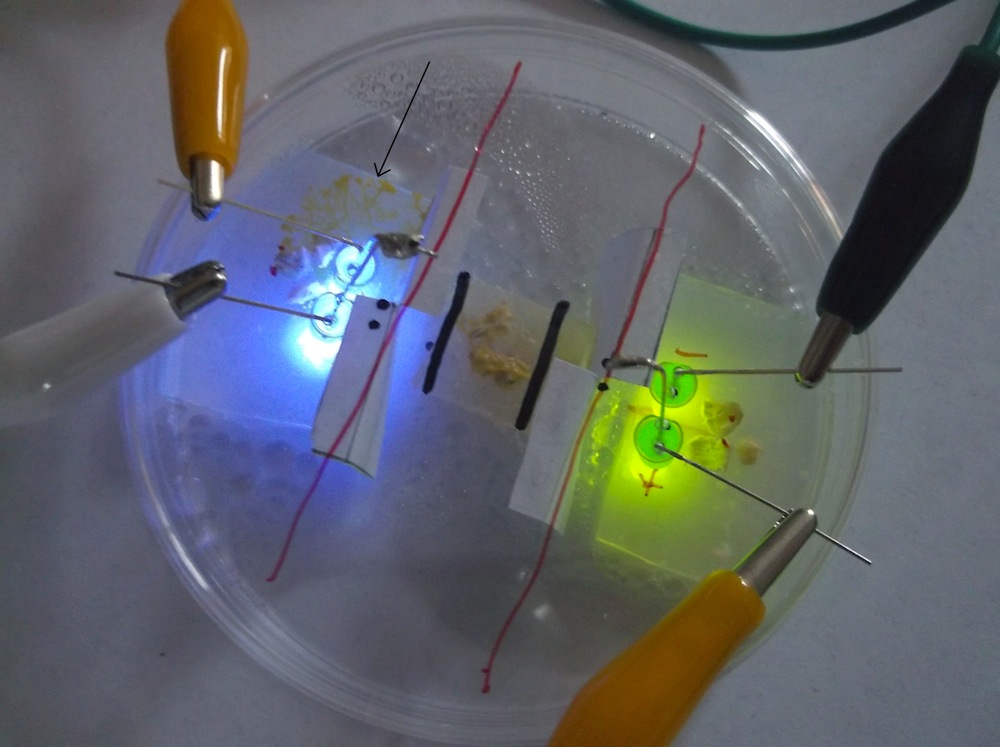}}
\caption{Phototaxis experiments. 
(a)~A scheme of experimental set-up for phototaxis experiments. Gray squares represent agar cubes: the slime mould is inoculated on the central cube (pentagon) and is left to proliferate. Black rectangles represent card barriers with a 0.5~mm gap underneath, which the plasmodium is able to migrate underneath to access the peripheral agar cubes, which are loaded with chemoattractants and illuminated by 2$\times$ LEDs (circles).
(b)~Photograph of a completed phototaxis experiment: note how the plasmodium has migrated towards the left 'blue' side (arrowed).}
\label{fig_taxis}
\end{figure}

Plasmodia were inoculated on to 2.5~cm\textsuperscript{2} squares of 2\% non-nutrient agar in the centre of 9~mm Petri dishes. Similar squares of agar loaded with chemoattractants (oat flakes) were situated at the 3 and 9 o'clock poles of the dishes  and were each illuminated by 2 lid-mounted LEDs (Fig.~\ref{fig_taxis}a). 4 colours of LED were used --- blue (466~nm), green (568~nm), yellow (585~nm) and red (626~nm) --- and each pole had a different colour assigned to it in each experiment. The peripheral regions were separated from the centre by a strip of thick card glued to the dish lid, with a 0.5~mm gap underneath: this prevented most of the light reaching the inoculation point, but allowed the plasmodium to migrate through the gap underneath. 

Test plasmodia were therefore given three basic choices in each experiment: migrate towards colour A, migrate towards colour B or do neither (and hence stay in the centre/sclerotinize/attempt to colonise the card/escape the dish etc.). The choices each plasmodium made were compiled into an order of preference for each colour: for each experiment, the colour which was avoided was given 1 'phobia point'; the colour with the most points after all experiments had been completed was determined to be the strongest repellent.

	A completed phototaxis experiment is shown in Fig.~\ref{fig_taxis}b. This pattern of avoidance was identical for every repetition. On no occasion did the plasmodium chose to stay on the central agar square and sclerotinize, although on two occasions it migrated up the card barrier and attempted to consume the glue securing it to the lid of the Petri dish. Phototaxis experiments also took considerably more time to complete than anticipated, with each lasting 5--6 days (whereas similar experiments with only chemoattractants/repellents typically last less time): the plasmodium would typically spend several days exploring the central agar piece and occasionally extend pseudopodia several millimetres along the plastic base of the Petri dish in random directions. This would seem to imply that the plasmodium's actions were a 'last resort' --- a protistic equivalent of reluctance.

\emph{P. polycephalum's} preference to each colour of LED is summarised below, arranged in order of most avoided $\rightarrow$ least avoided:
\begin{center}Green $\rightarrow$ Red $\rightarrow$ Yellow $\rightarrow$ Blue \\\end{center}

Thus green LEDs were used in further experiments.

\section{NOT Gate}

\begin{figure}[!tbp]
\centering
\subfigure[]{\includegraphics[width=0.9\textwidth]{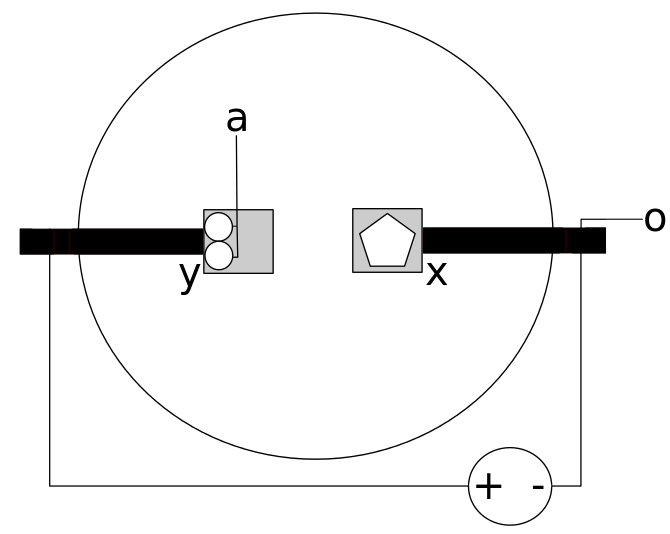}}
\subfigure[]{\includegraphics[width=0.49\textwidth]{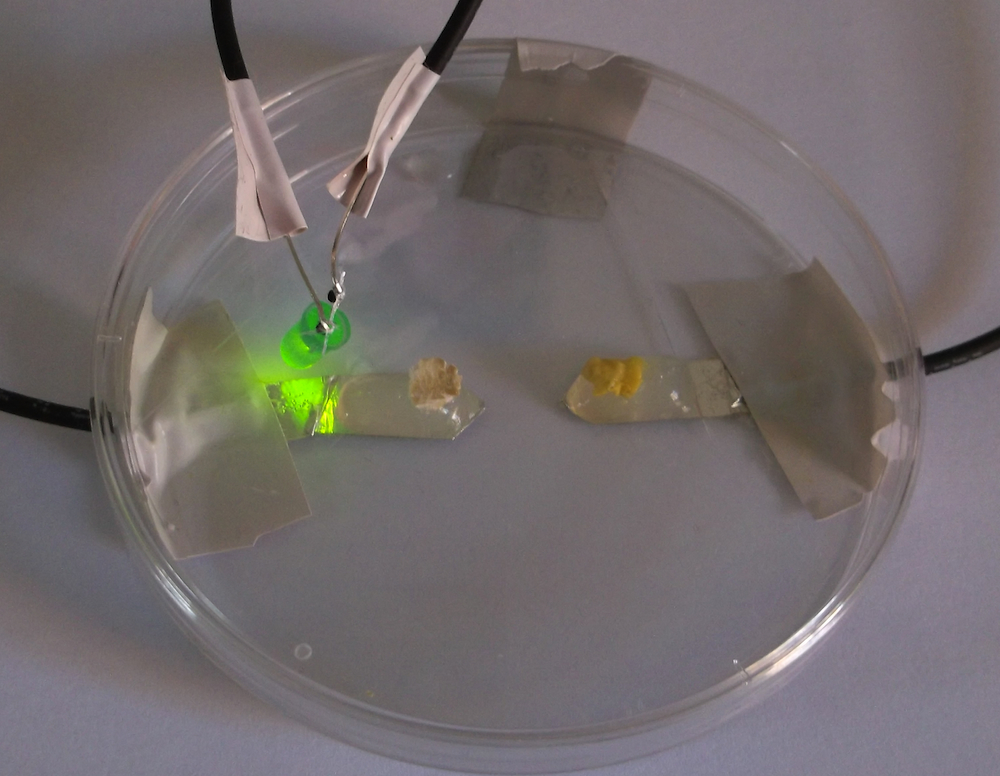}}
\subfigure[]{\includegraphics[width=0.49\textwidth]{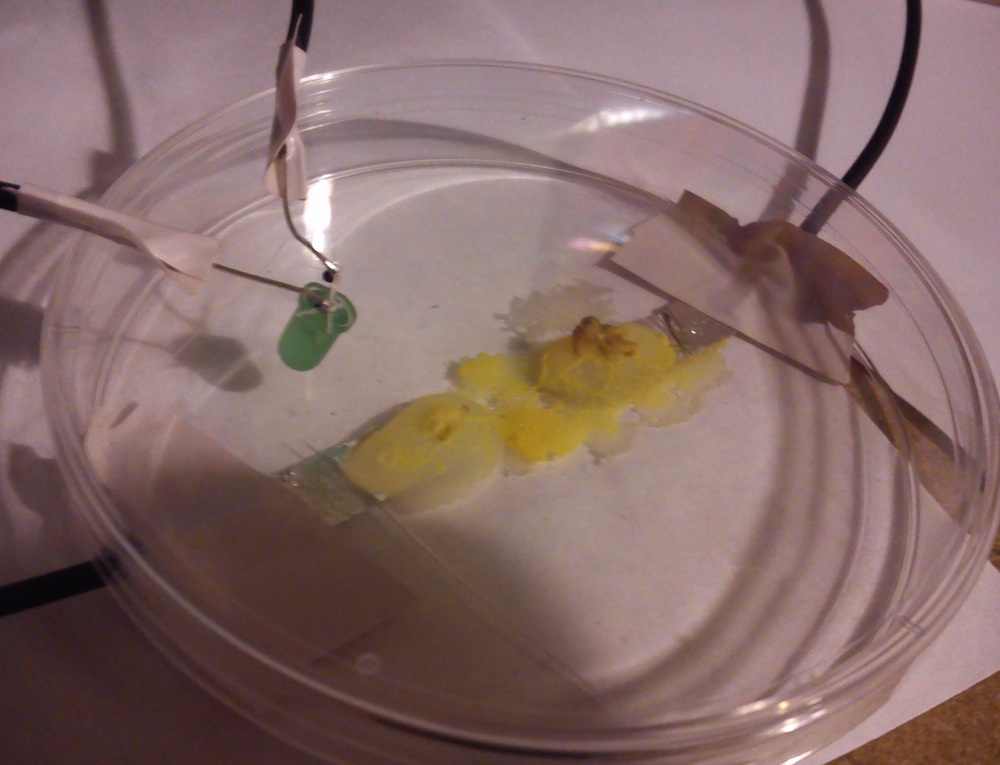}}
\caption{Physarum NOT logic gate.
(a)~Schematic diagram of a PNOT logic gate. A plasmodium (pentagon) is placed on agar (grey squares) overlying electrode X and chemoattractants are placed on Y. If the input [a] to the LEDs (circles)  is at logic 0, the LED does not switch on and a plasmodial tube is formed between electrodes X and Y; \emph{vice versa} also occurs. The electrodes are wired into an independently-powered output circuit [o].
(bc)~Photographs of a functional PNOT gate. 
(b)~Experimental setup with input at logic 1; the plasmodium did not migrate and the output was left at logic 0. 
(c)~With input at logic 0, the plasmodium has migrated across the gap and created a `wire' which was electrically conductive, leaving the output at logic 1.
}
\label{fig_pnot}
\end{figure}

	The design for the Physarum-NOT gate (PNOT) is shown in Fig.~\ref{fig_pnot}a. Two aluminium tape electrodes (X and Y) were stuck to the base of a 9~mm plastic Petri dish, separated by a central 10~mm gap. 2~ml agar blobs were placed on top of each electrode, one of which was loaded with plasmodium (X) and the other a chemoattractanct (an oat flake) (Y). The gate's 'input' consisted of a 9~V power supply (Lascar Electronics, Salisbury, UK) connected to a lid-mounted LED overlying electrode Y. At input equals 1 (logic levels of 0 and 1 are represented here as 0~V and 9~V, respectively), the LED was active and prevented the plasmodium from migrating from X to Y, preventing the completion of the circuit. If the input equals 0, the LED was not active and therefore the plasmodium was free to migrate to electrode Y, forming a continuous tube of plasmodium between both electrodes. 
	
	Both electrodes were connected to a separate 'output' circuit with a constant potential difference applied across it. The plasmodium, when spanning two electrodes, acted as a physical conductor of electricity and hence completed the output circuit. As such, the device's functionality is the same as a conventional NOT gate, i.e. if the input is live, the output is not and \emph{vice versa}. 
	
	Experimental photographs of a PNOT gate are shown in Fig.~\ref{fig_pnot}. The time it took for the plasmodium to complete its logic function was extremely variable, ranging from 1--4 days; approximately 25\% entirely failed to propagate. With reference to the results of the phototaxis experiments, this was reasoned to be due to the plasmodium's tendency to stay still for as long as possible before migrating, during which time the agar blobs may have desiccated to the point that they could no longer support its demand for moisture.

\section{NAND Gate}

\begin{figure}[!tbp]
\centering
\subfigure[]{\includegraphics[width=0.9\textwidth]{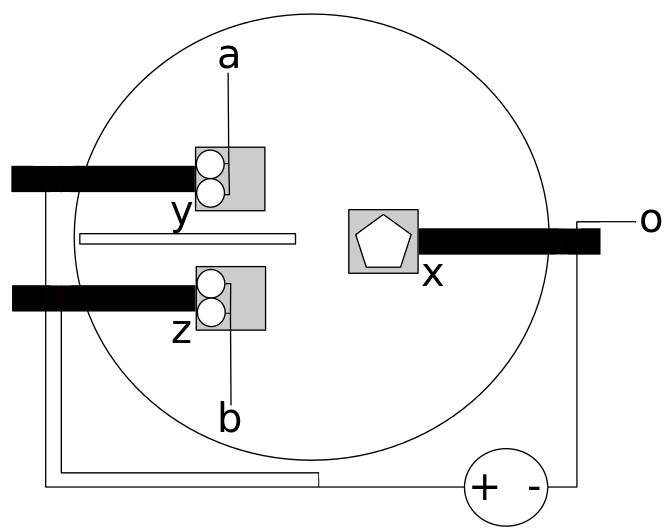}}
\subfigure[]{\includegraphics[width=0.49\textwidth]{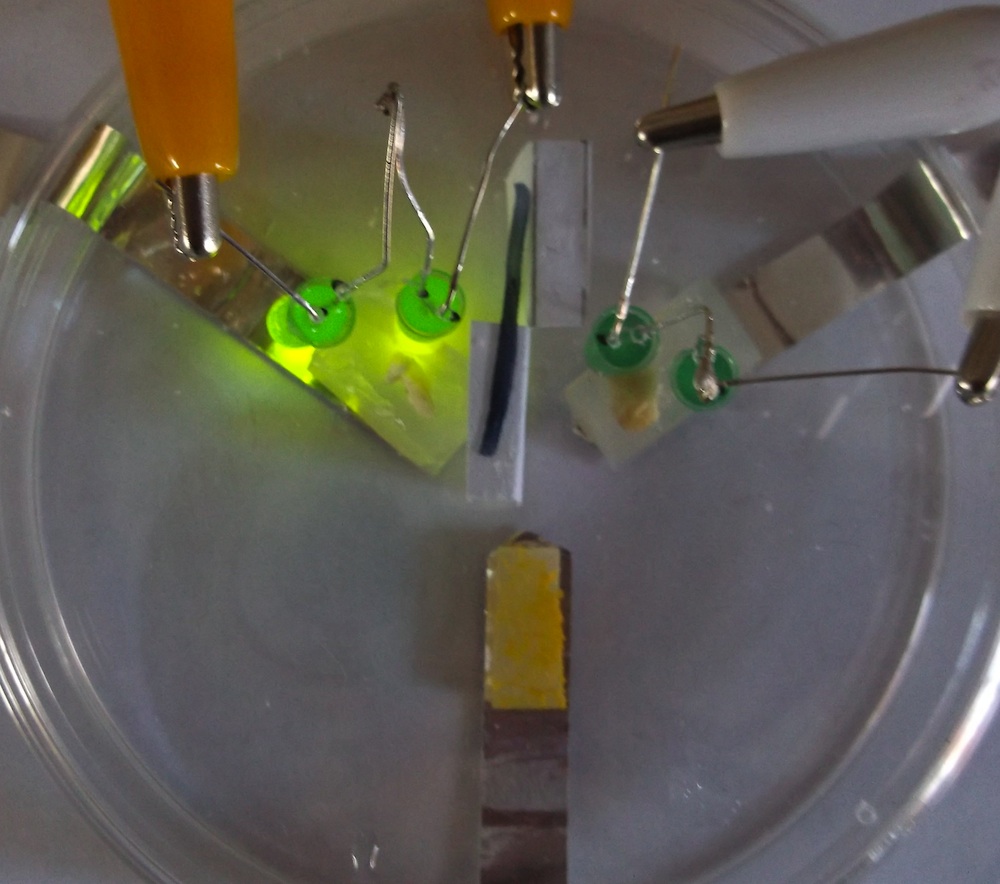}}
\subfigure[]{\includegraphics[width=0.49\textwidth]{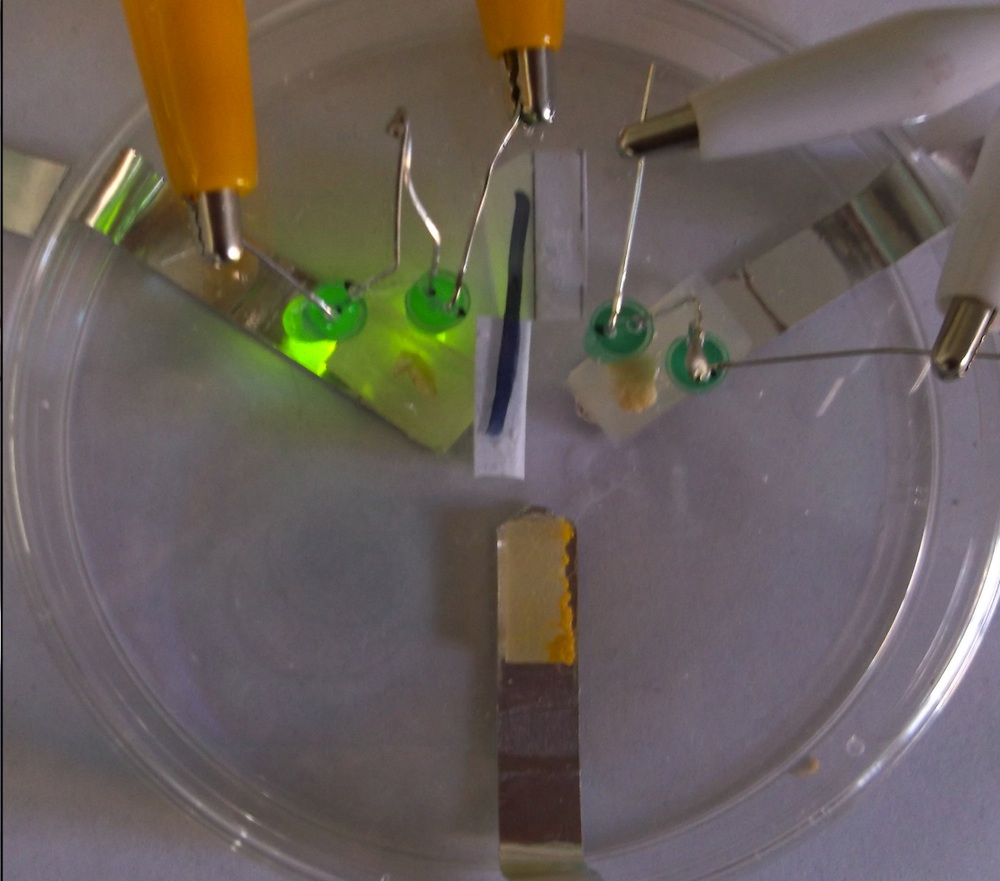}}
\subfigure[]{\includegraphics[width=0.49\textwidth]{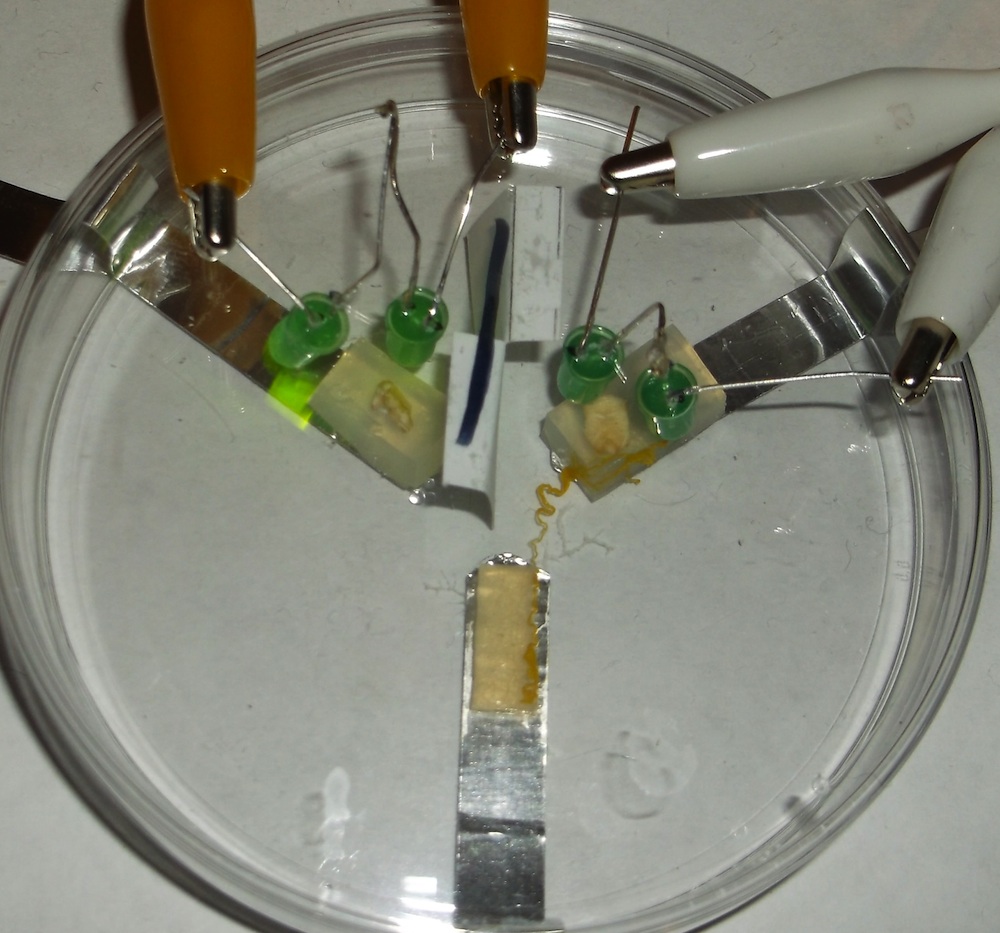}}
\caption{Physarum NAND gate.
(a)~Schematic diagram of a PNAND logic gate. Slime mould (pentagon) is placed on agar (grey squares) overlying electrode X and attractants are placed on Y and Z. If inputs A or B are at logic 1, the corresponding LEDs (circles) turn on, preventing slime mould growth to the corresponding electrode and hence stop a circuit from forming. If both A and B are on, no circuit forms, resulting in no output and \emph{vice versa}. The electrodes are wired into an independently-powered output circuit [o]. A card barrier separates electrodes Y and Z to prevent light contamination.
(bcd)~Photographs of a functional PNAND gate. 
(b)~T=6~h; the inoculation point is fully colonised and the left electrode is illuminated. 
(c)~ T=12~h; the plasmodium has migrated to the left of the inoculation point. 
(d)~T=24~h; the plasmodium has migrated to the unilluminated electrode.
}
\label{Fig_nand}
\end{figure}

Physarum-NAND (PNAND) gates were constructed based upon the same principles of the PNOT gate, but with a second electrode and LED added to the device (Fig.~\ref{Fig_nand}a). LEDs A and B were linked to separate power sources (i.e. inputs) so that their corresponding LEDs worked independently of each other. A plasmodium was once again loaded onto electrode X, but was given the added option of migrating towards electrode Z as well as Y. If both inputs equal 0, the plasmodium could migrate towards either Y or Z at random, forming a circuit between the initial and new electrode, causing the output  to equal 1. If input A is 1 but B is 0, the plasmodium at electrode X is still free to complete the circuit by migrating to electrode Z, and \emph{vice versa}. If both A and B equal 1, the plasmodium would not migrate towards either electrode, preventing a circuit from forming and hence leaving the output at 0. Electrodes X and Y were physically separated by a barrier similar to those described in Sec.~\ref{phototaxis} to prevent light contamination from opposing LEDs. Experimental photos of PNAND gate are shown in Fig.~\ref{Fig_nand}b--d.

\section{Reusability}

	Both PNOT and PNAND gates were tested for reusability. Once a logical operation had been carried out, the inputs were changed and the consequent responses of experimental plasmodia were recorded. 
	
Both PNOT and PNAND gates are reusable, but only a finite number of times. Resetting gates --- i.e. allowing the plasmodium to propagate between two unilluminated points then illuminating one of them in expectation that the plasmodium would withdraw --- was successful in that it caused the plasmodium to withdraw from the target electrode in all instances. This process usually took 2-6 hours and involved the contents of the plasmodium flowing back down the tubes spanning both electrodes. Occasionally, the plasmodium split into two separate fragments, both of which sclerotinized. With only two electrodes present, the PNOT gate was therefore effectively reset as the conductive trail between both electrodes was disrupted. PNOT gates were not successfully re-reset, i.e. the slime mould would not migrate back to an area it had already colonised and withdrawn from.

\begin{figure}[!tbp]
\centering
\subfigure[]{\includegraphics[width=0.47\textwidth]{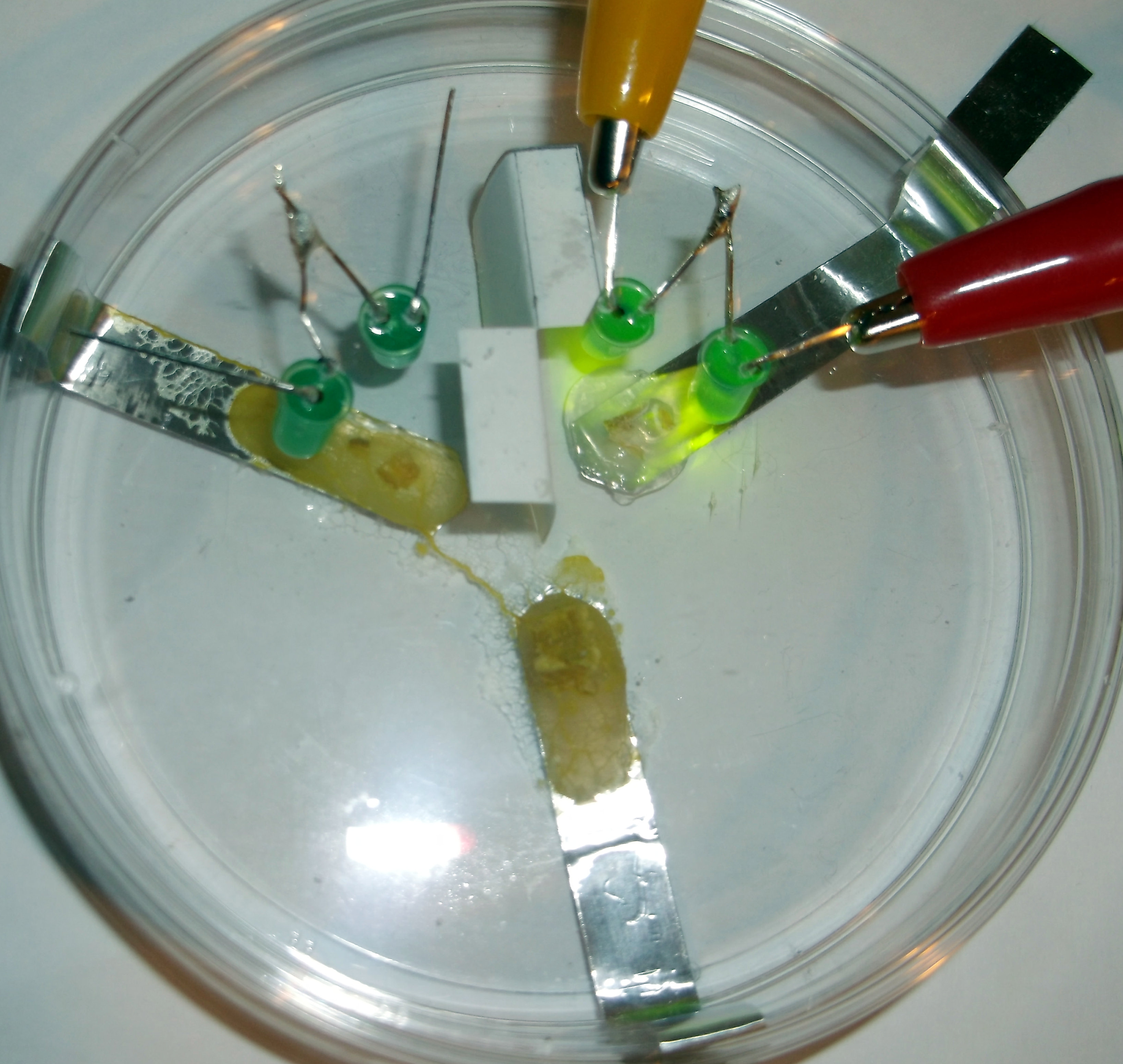}}
\subfigure[]{\includegraphics[width=0.51\textwidth]{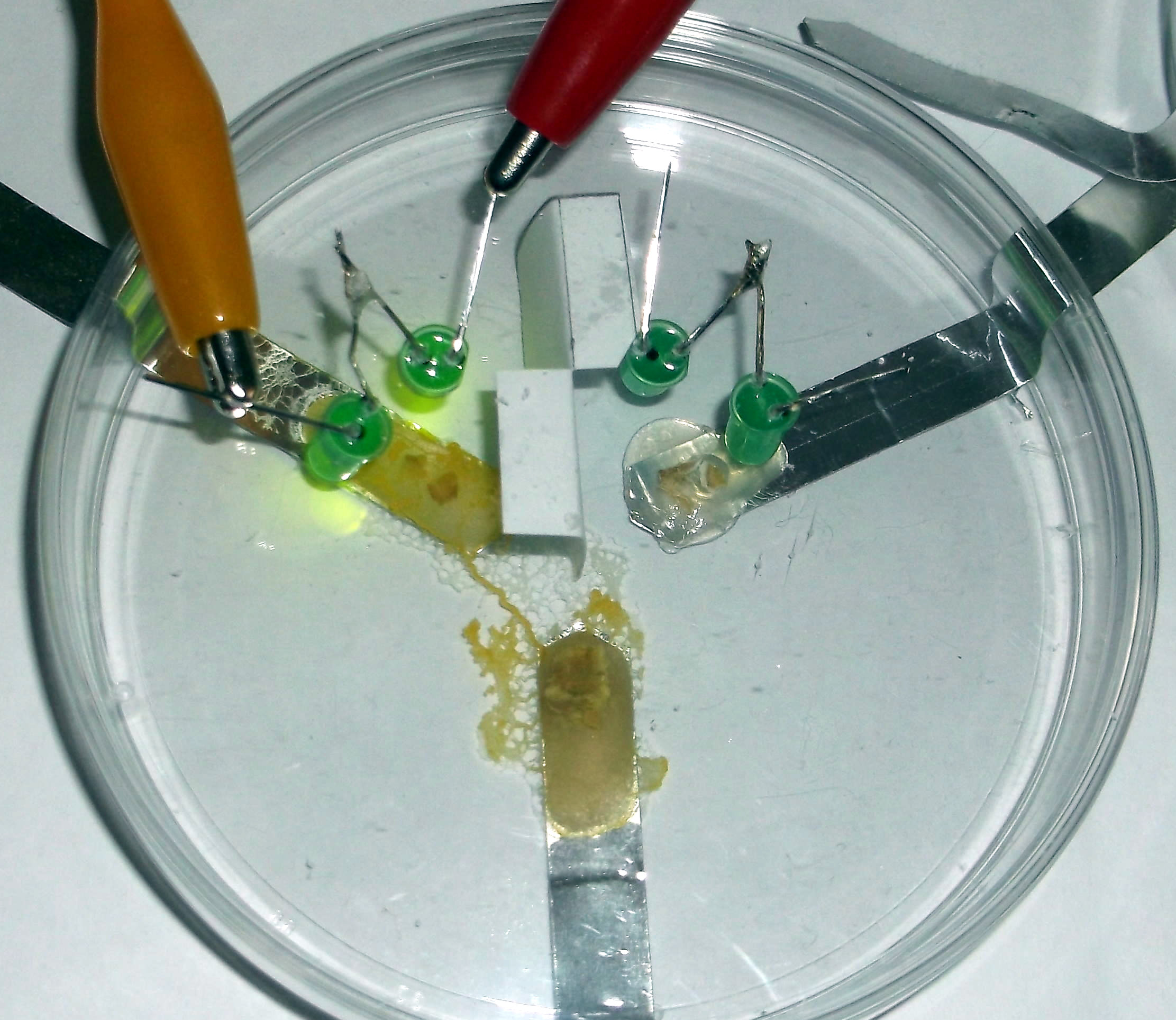}}
\subfigure[]{\includegraphics[width=0.51\textwidth]{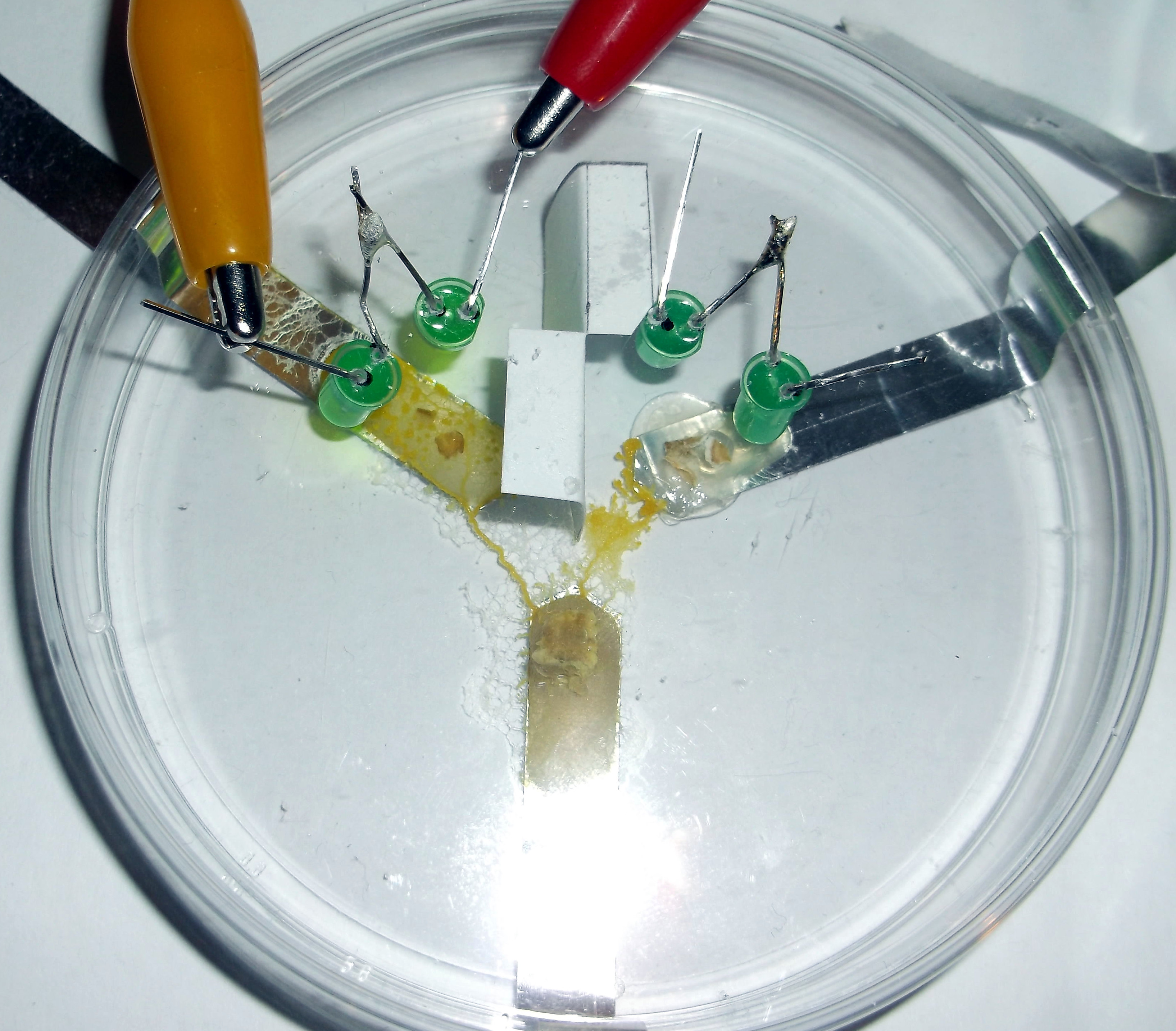}}
\caption{Resetting and reprogramming a PNAND gate.
(a)~T=36~h. A logical operation has been completed by the slime mould: the right LEDs are on and so it has migrated to the left electrode.
(b)~T=38~h. The inputs are switched; the right LEDs are now off and the left are on. The slime mould begins to withdraw towards the inoculation point and sends pseudopodia in multiple directions looking for a new place to migrate to.
(c)~T=42~h. The slime mould migrates towards the unilluminated right electrode and begins to withdraw its protoplasm from the left electrode and the tube between left and centre electrodes gradually ceases to be conductive.
}
\label{fig_reset}
\end{figure}

PNAND gates were successfully reset and reprogrammed, as shown in Fig.~\ref{fig_reset}: once a logical operation had been completed, the inputs were altered and the slime mould migrated accordingly to complete the new operation. PNAND gates that had been reset had a higher failure rate than `fresh' gates, but their propagation delay was significantly shorter (typically $<$~24 hours).

\section{Fault-tolerance}	
	
The fault tolerance of Physarum gates was also investigated by altering physical variables in the test environment, including:
\begin{enumerate}
\item Increasing the luminosity of the LEDs by replacing them with brighter variants (each was increased by approximately 50~mcd).
\item Increasing the distance between electrodes to 20~mm
\item Increasing the potential difference between electrodes to 24~V
\end{enumerate}

None of these variables were `decreased', as the initial variables were originally designed as minimal operational parameters; dimmer light sources were anticipated to be ignored by the plasmodium, decreasing the electrode-electrode gap to below 10~mm could result in the plasmodium easily spanning all three (resulting in gate mis-operation) and applying a lower voltage through the electrodes would likely have resulted in an output so low as to be undetectable.
	
	Whilst adjusting these variables is of little practical use in the design of these specific prototype devices, they were chosen to (a) investigate whether fine-tuning any of these variables would have a beneficial effect on the operation of the device, and (b) demonstrate how a Physarum machine may adapt to unfavourable conditions in a way that traditional computing architectures cannot --- e.g. it is doubtful that any solid state architectures would function properly following exposure to prolonged overvoltage. By introducing extra elements of environmental instability, \emph{P. polycephalum} was given the opportunity to demonstrate how resilient it is, and therefore how robust a Physarum machine could potentially be.
	 	 
Increasing the luminosity of the LEDs appeared to have no beneficial or detrimental effect on their operation; failure rates were not significantly different and neither was the operation time/propagation delay. Whether \emph{P. polycephalum} is generally phobic to any intensity of green light, or whether the increase in incandescence was too low to provoke any stronger responses was not ascertained.

Doubling the distance between the electrodes markedly increased the failure rate of both varieties of Physarum logic gate. This was likely to result from the combined effects of \emph{P. polycephalum's} reluctance to cross the arid plastic regions of Petri dishes and a lack of any detectable nutrient gradients due to the distant proximity of the supplied chemoattractants.

Experimental plasmodia appeared to be extremely tolerant of high voltages. Whilst the current that would have run through the plasmodium would have been comparatively low due to the resistance of the agar blobs and the innate resistance of the plasmodium its self, it was nevertheless tolerant to applied potential differences of up to and including 24~V. The morphology of plasmodia exposed to high voltages was slightly different to the wild type; rather than spanning the gap between electrodes with a single tube of plasmodium, plasmodia exposed to higher voltages tended to occupy a significant amount of space between electrodes with multiple interlinking tubules (Fig.`\ref{Fig_5}b).  

\section{Conclusions}

The principle finding of this study was that it is indeed possible to `program' an unaltered \emph{P. polycephalum} plasmodium to carry out basic logic functions with optical inputs. Whilst the prototype devices presented here have significant limitations, they demonstrate a genuine implementation of unconventional computation in which a living computing substrate functions as both a sensing, decision making computing device and a physical conductor of electricity.

\emph{P. polycephalum's} photoavoidance is widely mentioned in literature, and indeed the plasmodium is best cultivated in the dark. Sauer~\cite{Sauer_1982}  reports that the plasmodium is phobic to blue light (which may also induce sporulation whilst the plasmodium is also starved of nutrients and/or moisture), but may sometimes migrate towards red light. Green light has been reported to inhibit sporulation, and seemingly random wavelengths of every colour (and some UV wavelengths) induce a range of metabolic changes \cite{Hato_etal_1976,Sauer_1982}. Whilst \emph{P. polycephalum} is not photosynthetic, it contains pigments similar to phytochromes which are thought to act as photoreceptors \cite{Kakiuchi_etal_2001}. A migrating plasmodium may be diverted by placing illuminated barriers in its path \cite{Adamatzky_2009}.

There were clear discrepancies between the results of the phototaxis experiments conducted for this investigation and those in published literature. There were, however, stark differences in experimental setup between the experiments described here and those in literature; LED diffusion, different wavelengths of light used, different intensities of light and different experimental conditions could all have altered plasmodial responses drastically. It should also be noted that if, for example, the plasmodium migrated towards blue light, this does not necessarily imply that \emph{P. polycephalum} is preferentially moving towards the blue light, but is more likely moving towards a nutrient source out of necessity and can tolerate blue light better than the alternative. Functionally, however, these distinctions are unimportant as the most-avoided colour of light is the most appropriate for use as a repellent. LEDs were chosen for this investigation for being cheap, utilitarian, small and power efficient: these are all highly desirable characteristics in emergent computing devices.

The most significant limitation of this model was the vast propagation delays which prevent the construction of combinational logic circuits, e.g. adders. The prototype gates produced here therefore hold no specific practical value other than demonstrating an implementation of optically-programmable logic. Aside from optimising the experimental environment (reducing the effects of desiccation, changing the size of the gaps between electrodes etc.), the most viable research direction for producing gates of practical value would appear to be minimising the role of physical movement. In Ref. \cite{Adamatzky_Jones_2011}, Adamatzky and Jones demonstrated that \emph{P. polycephalum} responds to a range of environmental stimuli with predictable patterns of electrical activity (when measured at the outer membrane); if the electrical responses of the plasmodium to light are similarly predictable, culturing the plasmodium on a large array of electrically conductive surfaces would therefore allow the researcher to track plasmodial responses electrically. This could potentially vastly reduce propagation delays and hence allow for gate cascading.
	
Another limitation of this model was physical size. The test environments used (Petri dishes) were 9~cm in diameter, meaning that a single gate was comparatively large, especially when including the breadboard and power supply. If the aforementioned propagation delays could be negated, combinational logic circuits would still be large and ungainly --- e.g. a one-bit half adder comprised of 7 PNAND gates would use approximately 0.5~m\textsuperscript{2} on a bench top --- and therefore of limited practical value. If future prototypes still relied on slime mould taxis, each test environments could theoretically be scaled down to approximately 3~cm in diameter. Multi-gate testing environments with in-built mechanisms for cascading gates and reducing solid-state components could further scale the devices down.
	
	The electrical output of these gates was extremely low, even when the output circuit voltage was increased to over 24 Volts. This was likely in part due to the high resistance created by the agar blobs (which were measured in Ref. \cite{Mayne_Adamatzky_2013} and found to have an average resistance of 18K$\Omega$ each); any next-generation device would therefore benefit from using a more electrically-conductive substrate. This could present significant barriers to research as conductive gels, such as those used for skin-electrode contacts in medical science, typically contain high amounts of salt to facilitate electrical conductivity which would likely be harmful to the plasmodium.
		
		After discussing the limitations and failings of these devices, it is pertinent to mention their redeeming features. The devices produced here were all cheap and comparatively easy to fabricate; they are reusable and produce no hazardous waste. The power consumption was also relatively low, although the power to processing power ratio was far inferior to conventional solid-state hardware. Furthermore, the apparent robustness of slime mould devices to adverse conditions such as overvoltage highlights how Physarum-based computing may be adapted to roles to which standard computing architectures are not suited, e.g. in harsh conditions applications and research. The experiments designed here may also be used to model and approximate the growth and behaviour of other similarly structured (i.e. branching) cells or organisms, such as neurons, melanocytes, plant roots and fungi.
	
	It is relevant to note that although the devices presented here principally relied on light as a controlling input, other inputs are entirely feasible to explore. For example, \emph{P. polycephalum} also exhibits magnetotaxis \cite{Pazur_Schimek_Galland_2007}, thermotaxis (to an ideal temperature of 29$^\circ$C) \cite{Tso_Mansour_1975}, negative chemotaxis away from repellents \cite{Costello_Adamatzky_2013} and negative thigmotaxis \cite{Adamatzky_2013b}, to name a few of the other senses the plasmodium is able to perceive and coordinate a response to. Any of these may be used as viable attractants/repellents in slime mould devices. Indeed, a putative next-generation slime mould logic device could employ multiple inputs.
	
	Although revisiting moving-part computing elements is essentially retrogressive, it must be remembered that a Physarum machine is not a linear device, i.e. its actions are not comparable to, for example, a vacuum tube or other moving component whose functionality is extremely limited. The Physarum machine's genetically-encoded behaviour - analagous to firmware - bestow it with a form of judgement which is expressed in its highly efficient foraging behaviour. If a computer is given a command with one figure wrong, the program will terminate; the Physarum machine will adapt and dynamically react to any range of inputs it is given. This ability to reason (automated reasoning) is not a property which has been fully achieved with solid-state hardware.\\

\end{document}